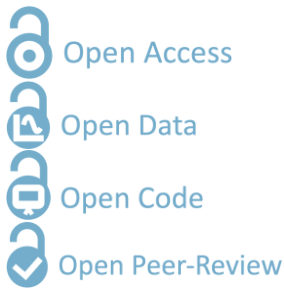

RESEARCH ARTICLE

# New insights into the population genetics of partially clonal organisms: when seagrass data meet theoretical expectations

Arnaud-Haond, Sophie[1]; Stoeckel, Solenn[2]; Bailleul, Diane[1]

[1]MARBEC, Ifremer, Laboratory Environnements-Ressources, Ifremer, Bd Jean Monnet BP 171, Sète 34203, France.
[2]INRA, UMR1349 Institute for Genetics, Environment and Plant Protection, Le Rheu, France.

This article has been peer-reviewed and recommended by
*Peer Community in Ecology*

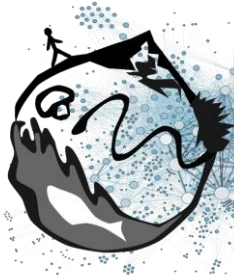

**ABSTRACT**

Seagrass meadows are among the most important coastal ecosystems, in terms of both spatial extent and ecosystem services, but they are also declining worldwide. Understanding the drivers of seagrass meadow dynamics is essential for designing sound management, conservation, and restoration strategies. However, the poor knowledge of the effect of clonality on the population genetics of natural populations severely limits our understanding of the dynamics and connectivity of meadows. Recent modelling approaches have described the expected distributions of genotypic and genetic descriptors under increasing clonal rates. Here, in light of these recent theoretical developments, we revisited population genetic data for 165 meadows of four seagrass species. Contrasting shoot life span and rhizome turnover led to the prediction that the influence of asexual reproduction will increase along a gradient from *Zostera noltii* to *Zostera marina*, *Cymodocea nodosa* and *Posidonia oceanica*, which should be reflected by an increasing departure from Hardy-Weinberg equilibrium ($F_{is}$) and decreasing genotypic diversity ($R$). This meta-analysis provides a nested validation of this hypothesis at both the species and meadow scales through a significant relationship between $F_{is}$ and $R$ within each species. By empirically demonstrating the theoretical expectations derived from recent modelling approaches, this work calls for the use of Hardy-Weinberg equilibrium ($F_{is}$) rather than the strongly sampling-sensitive genotypic index $R$ to assess the importance of clonal reproduction, at least when the impact of selfing on $F_{is}$ can be neglected. The results also emphasize the need to revise our appraisal of the extent of clonality and its influence on the dynamics, connectivity and evolutionary trajectory of partial asexuals in general, including in seagrass meadows, to develop the most accurate management strategies.

*Keywords:* seagrass, mating system, clonal growth, dispersal, marine meadows

## Introduction

Clonality is a life history trait spread across the Tree of Life (Halkett et al., 2005) that characterizes the species forming the basis of most important terrestrial and marine ecosystems. The drastic decline in many ecosystems engineered by partially clonal species (Carlsson Callaghan, 1994; Carpenter et al., 2008; Polidoro et al., 2010; Waycott et al., 2009) emphasizes the need to identify the drivers and life history traits, including reproductive strategies, underlying demographic declines or the colonization of new areas during range collapses, shifts, or expansions (Aitken et al., 2008; Callaghan et al., 1992; Cornelissen et al., 2014; Pecuchet et al., 2018; Yu et al., 2016). Although population genetics can provide essential indirect evidence, the use of population genetics has been limited thus far by the lack of clear theoretical predictions on the effect of partial asexuality on the distribution of genetic polymorphism, hampering the comprehensive analysis and interpretation of population genetics data.

The use of molecular markers has led to major improvements in the study of the architecture, dynamics, and evolution of clonal organisms, particularly since the development of molecular and analytical methods to detect clonality (Halkett et al., 2005; Tibayrenc et al., 1990) and to assess clonal membership (Arnaud-Haond et al., 2007b; Tibayrenc et al., 1990). These improvements have allowed the identification in natural populations of clonal replicates (i.e., ramets) of distinct "genetic individuals" (i.e., genets) through the recognition of clonal lineages (multi-locus genotypes, MLGs, or multi-locus lineages, MLLs). This has permitted the estimation of not only genetic (i.e., allelic richness $A$ and heterozygosity $H$) but also genotypic (i.e., indices based on the number of MLGs or MLLs) diversity.

The simplest index of genotypic richness, $G$, corresponds to the number of MLGs or MLLs (Arnaud-Haond et al., 2007b). Since this specific unit of evolution (i.e., the genotype: Ayala, 1998) became accessible, genotypic diversity has been the subject of a growing number of studies aiming to investigate the resistance of natural populations in diverse environmental conditions (Hughes et al., 2008; Massa et al., 2013; Reusch et al., 2005). Such information is essential for understanding the evolution and dynamics of natural populations, including populations of environmental engineers (Callaghan et al., 1992; Cornelissen et al., 2014), such as corals and seagrasses, that form the basis of essential and declining coastal ecosystems (Carpenter et al., 2008; Hughes Stachowicz, 2009; Orth et al., 2006). $G$ naturally increases with the sample size in terms of the number of ramets (Ellstrand & Roose 1987; Dorken & Eckert 2001). A variety of richness and diversity metric indices have thus been used to describe clonality in natural populations; some of these indices, such as the Shannon and Simpson indices, were borrowed from biodiversity literature, while others were simple indices ($Pd = G / N$ and $R = (G-1) / (N-1)$) based on the ratios of different genotypes ($G$) to the sample size ($N$) (Arnaud-Haond et al., 2007b). However, the versatile and inconsistent use of these indices, combined with a lack of a common standardized sampling strategy, have prevented sound biological comparisons of the extent of clonal reproduction and its consequences on the ecology and evolution of the diversity of studied organisms (Arnaud-Haond et al., 2007a). For approximately a decade, $R$ has been more consistently and widely used than other indices. Initially, thought to allow comparative studies, $R$ is also sometimes considered a *proxy* for the relative influence of sexual *versus* clonal reproduction and the consequences of such a reproductive strategy for the dynamics of natural populations, the mutation models, the main unit targeted by natural selection and drift (genotypes or alleles), and the main drivers of migrations (propagules or adults/fragments for plants, fungi and several invertebrates such as corals). Unfortunately, the strong dependence of $R$ on sampling strategy and density has also been clearly demonstrated in two previous studies (Arnaud-Haond et al., 2007c; Gorospe et al., 2015). In fact, those two studies, one based on a subsampling approach applied to two seagrass species and the other based on the exhaustive genotyping of a coral reef, showed a complete lack of stabilization of *R* with increasing sampling size and density. Basically, the index declines as new samples are added. This unfortunate property results in an inability to derive any equivalence between the *R* value and the extent of clonal reproduction, *c*, thus jeopardizing the management and conservation strategies for populations of partially asexual organisms and the ecosystems these populations support. In fact, only two recent programmes allow the computation of *c* from genetic data, and their use relies on the important but seldomly fulfilled requirement of temporal samples; these programmes are CloNcaSe (Ali et al., 2016), a method used for organisms with cyclical parthenogenesis, and ClonEstiMate (Becheler et al., 2017), a Bayesian method for samplings ideally separated by one generation.

As for genetic diversity indices, they are seldom used in ecological studies to appraise the influence of clonality. This may be partly because pioneer mathematical models suggest that the clonal rate has a very limited influence on the genetic composition of populations reported through $F_{is}$ (de Meeûs et al., 2007) and linkage disequilibrium (Navascues et al., 2010) indices. These models suggest that a departure from Hardy-Weinberg equilibrium (HWE) towards heterozygote excess and linkage disequilibrium would be a signature of nearly exclusive clonality (Halkett et al., 2005). However, almost exclusively clonal lineages are exceptions, resulting in the interpretation of negative $F_{is}$ values (heterozygous excess), when not overlooked or hastily discarded through over-conservative corrections for multiple tests (Bonferroni correction, Rice, 1989), as indicative of high clonal rates (Halkett et al., 2005). Such a path has been followed to interpret heterozygote excess in organisms such as pea aphids (Delmotte et al., 2002; Halkett et al., 2005b) and cultivated algae (Sousa et al., 1998; Guillemin et al., 2008) as a signature of clonal dominance, while the lack of a departure from HWE, together with a high *R*, has been interpreted as indicative of predominant sexual reproduction (Krueger-Hadfield et al., 2011). More recent mathematical models have suggested the occurrence, at equilibrium, of larger inter-locus variance in $F_{is}$ values and an expected departure from HWE as the rate of clonality increases (Stoeckel Masson, 2014), even for modest rates of clonality (Reichel et al., 2016; Stoeckel et al., 2019). These findings were more recently used to enlarge the interpretation of deviations from HWE as indicative of partial (though not necessarily extreme) clonality in organisms such as invasive algae (Krueger-Hadfield et al., 2017)

Taken together, the severe sampling bias associated with *R* and the expected influence of partial clonality on genetic parameters may explain the paradoxical observation of heterozygote excess in partially clonal organisms (Reichel et al., 2016), despite the often elevated values for clonal richness, which are possibly partly due to a low sampling density (Arnaud-Haond et al., 2007c; Gorospe et al., 2015). Seagrasses are a good example of this paradox. On the one hand, moderate to high levels of clonal richness, *R*, together with limited but significant values of genetic differentiation ($F_{st}$), whether at the ramet or at the genet level, have led some authors to propose a strong influence of sexual reproduction on the dynamics and evolution of seagrass meadows, implying important recombination rates and large-scale dispersal (Kendrick et al., 2012; McMahon et al., 2014). On the other hand, the maximum values that $F_{st}$ can reach are strongly limited by the use of highly polymorphic molecular markers (Hedrick, 1999a, 2005; Jost, 2008). In addition, elevated values for clonal richness often appear with significant but often-overlooked heterozygote excesses, thus paradoxically suggesting a high incidence of clonality (Reichel et al., 2016; Stoeckel et al., 2019).

To resolve this paradox and better understand the meaning of *R* and *F* values as they relate to the extent of clonal reproduction and the importance of dispersal, we propose a re-analysis of previously published seagrass data in light of recent modelling developments describing the effect of increasing *c* values on the genotypic and genetic descriptors (including *R* and $F_{is}$) of populations (Stoeckel et al., 2019). These four seagrasses are the main species structuring coastal ecosystems along the Atlantic and Mediterranean coasts and exhibit a gradient from long-lived species exhibiting a slow turnover of shoots and low to moderate reproductive output, namely, *Posidonia oceanica* and *Cymodocea nodosa,* to shorter-lived species, namely, *Zostera marina* and *Zostera noltii*, which exhibit an increased shoot turnover and allocation to sexual reproduction (see Box 1, Methods section and Table 1). Considering the expected influence of clonality based on the knowledge of clonal growth and the allocation of the four seagrass species and more recent theoretical predictions of increasing heterozygote excess with increasing levels of clonality, we expected a progressive increase in clonal diversity and a progressive decrease in heterozygosity along this gradient from *P. oceanica* to *Z. noltii*, starting with negative $F_{is}$ values (heterozygote excess) and moving towards null (or slightly positive when heterozygote deficiency occurs due to inbreeding) $F_{is}$ values for the shortest-lived species (Box 1).

Here, we aimed to test these theoretical expectations by using a meta-analysis of population genetics data published on these four seagrass species to answer the following questions:

1) Do estimates of genotypic diversity deliver informative values that can be used to assess comparative investments in clonal *versus* sexual reproduction among species submitted to a similar sampling scheme?

2) Does departure from HWE in natural meadows reliably reflect the relative importance of clonal *versus* sexual reproduction among species with different investments in clonal growth and different life spans?

3) Do the two families of parameters (genotypic and genetic) provide congruent qualitative estimates at both the interspecific and intraspecific levels?

**Box 1:** Main biological features of the four studied seagrasses and the associated hypothesis about the clonal architecture and genetic signatures.

Studies of the growth and dynamics of the four main European seagrasses, which inhabit coastal habitats from the inter- and subtidal compartments (*Z. marina* and *Z. noltii*) to water depths reaching depths up to and occasionally greater than 50 m (*C. nodosa* and *P. oceanica*), have shown marked discrepancies in the shoot growth, turnover and life span of these species, as well as investments in sexual reproduction. This implies differences in the relative incidence of clonal reproduction in the short and long terms, affecting the pattern of space occupation, as well as the temporal dynamics of shoots and meadows at both ecological and evolutionary time scales. These are summarized in the scheme below, together with a hypothesis about the expected clonal signature tested in the present work and parameters describing the clonal richness (*R*) and genetic composition ($F_{is}$) of meadows.

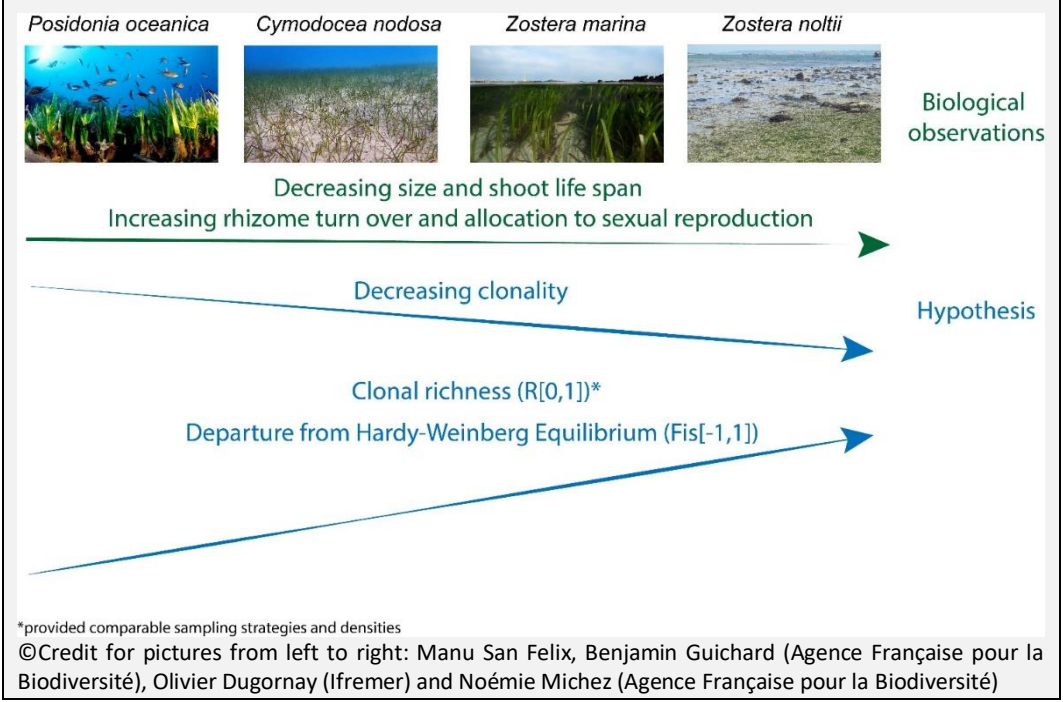

©Credit for pictures from left to right: Manu San Felix, Benjamin Guichard (Agence Française pour la Biodiversité), Olivier Dugornay (Ifremer) and Noémie Michez (Agence Française pour la Biodiversité)

## Methods

*Studied species*

Seagrasses reproduce clonally through rhizome elongation, and the rhythm and shape of this process differ among the studied species and largely scale with plant size (Duarte, 1991a; Marba Duarte, 1998). *Posidonia oceanica* and *Cymodocea nodosa* are known to exhibit more stable rhizome connections, a longer shoot life span and a lower shoot turnover than the other species (Diaz-Almela et al., 2008; Duarte, 1991a; Marba et al., 1996; Marba et al., 1998). In contrast, *Zostera* species exhibit the fastest growth but a more limited occupation of space (Marba et al., 1998; Sintes et al., 2006), as well as a much shorter life span.

Plant size thus appears to be strongly related to turnover time and module (i.e., rhizome internodes, leaf clusters and roots) longevity (Duarte, 1991a). These parameters of clonal growth are accompanied by a relatively parallel trade-off between clonal architecture and sexual reproductive output. The events and success of flowering are highly temporally and spatially heterogeneous for *P. oceanica* and *C. nodosa*, which seldom produce more than several tens of seeds per square metre; in contrast, highly profuse episodes of annual flowering in *Zostera* can lead to the production of thousands of seeds per square metre (Marba et al., 1996; Marba et al., 2004). Finally, both *Zostera sp*. and *P. oceanica* are hermaphrodites, while *C. nodosa* is dioecious (Larkum & Den Hartog, 1989).

These descriptions of clonal architecture and sexual reproduction features allow us to expect a decreasing influence of clonality from *P. oceanica* to *C. nodosa*, *Z. marina* and *Z. noltii* (see Table 1). Although no numerical estimates of shoot life span and rhizome turnover are available for the smallest and shorter-lived species *Z. noltii*, it is expected to exhibit a much higher turnover than the other species, in line with its habitat, which is characterized by relatively unstable environmental conditions (Duarte, 1991b). The fact that *C. nodosa* is an obligate outcrosser while other species can theoretically practice selfing may however result in different effects of sexual reproduction and clonality on the distribution of heterozygosity ($F_{is}$) in this species compared to the effects on the other three and is likely to affect the gradient expected on the basis of the sole influence of clonality (Box 1).

Interestingly, these species also often exhibit a gradient on the shore (Duarte, 1991b). *P. oceanica* and *C. nodosa* reach depths up to 40 to 50 metres (with *P. oceanica* dominating at greater depths, den Hartog, 1970), while the two *Zostera* species are most often encountered in intertidal areas, with *Z. noltii* sometimes exposed to desiccation and stressful conditions (Massa et al., 2009). All four species have been the focus of large-scale genetic surveys in phylogeographic studies in the past few decades. Datasets are thus available that can be used to test expectations about the investment of the species in clonal reproduction and to screen for signatures of effects of clonal reproduction on their genetic composition.

*Genetic datasets*

The *P. oceanica* dataset contains 36 samples of approximately 40 units each, with a total of 1424 shoots or connected shoots representing sampling units (SUs) ranging from Spain to Cyprus, a range of over 4000 km of coastline. For each sampling site, the SUs were collected according to randomly defined coordinates in a quadrat 80 metres long and 20 metres wide. Seven microsatellite markers were used: Po15, Po5, Po5-40, Po5-49, Po5-10, Po4-3 and Po5-39. The database of genotypes was constructed in a previous meta-analysis (Arnaud-Haond et al., 2014, Arnaud-Haond et al., 2014b for the dataset). Three meadows that were initially studied for the impact of farms on the genetic composition of the species (the populations Amathous, Acqua Azzura and Agias Nicholaos; Diaz Almela et al., 2007) were discarded to avoid introducing bias.

The *C. nodosa* dataset (Alberto et al., 2008) consisted of 47 meadows containing approximately forty SUs each, with a total of 1586 SUs from Cyprus to the Canary Islands and Madeira. For each meadow, the SUs were selected randomly in a quadrat 60 metres long by 14 metres wide. Eight microsatellite markers were used: Cn2-38, Cn2-14, Cn2-24, Cn4-19, Cn2-16, Cn2-18, Cn4-29 and Cn2-45 (Arnaud-Haond et al., 2014; Arnaud-Haond et al., 2014b for the dataset).

The first dataset for *Z. marina* consisted of 13 quadrats sampled in 7 meadows with 30 SUs per quadrat, resulting in a total of 390 SUs. The meadows were on the Brittany coast from Saint-Malo to Arradon. For each sampling site, SUs were randomly selected in two quadrats 20 metres long by 30 metres wide in 2009 (Becheler et al., 2013; the same sampling was performed in 2011, but in order to avoid partially duplicate information, only the first time series was included). A total of 9 microsatellite markers were used: GA35, GA2, GA17H, GA23, GA12, GA19, GA20, GA16 and GA17D (Becheler et al., 2014; Becheler et al., 2010; Becheler et al., 2013 for the dataset). The second dataset consisted of 14 meadows from Greenland to Southern Iberia, in which 427 SUs at least 1–1.5 m apart were collected haphazardly (Diekmann Serrão, 2012a) and genotyped with 8 microsatellite markers: GA6, GA3, CT17H, CT19, CT3, CT20, GA2 and CT35 (Diekmann Serrão, 2012b). This second dataset was not available and was thus used only for the analysis of $R$ and $F_{is}$ without replicates, as disclosed in the aforementioned article.

The *Z. noltii* dataset included 33 meadows sampled across the entire geographic range of the species and genotyped with 9 microsatellite markers: ZnB1, ZnH10, ZnB3, ZnB8, ZnH8, ZnD6, ZnE7, ZnF8 and ZnF11 (Coyer et al., 2004). Different sampling protocols were used and are detailed in the original publication; these protocols include the random collection of SUs within quadrats 60 metres long by 20 metres wide, 10 metres long by 15 metres wide and 10 metres long by 25 metres wide, as well as sampling along a linear transect every metre. One sampling site (BSea3) represented by a single SU was excluded from the initial dataset; thus, 32 meadows were kept for a total of 1117 SUs.

*Genetic and genotypic indices*

RClone 1.0.2 (Bailleul et al., 2016) in *R* statistical software (R Core Team, 2015) and GenClone (Arnaud-Haond Belkhir, 2007a) were used to manage the different datasets and to compute the genetic and genotypic indices of interest. We chose to examine the $F_{is}$ genetic index, for which models delivered several theoretical predictions, and the genotypic clonal richness index *R*, often used to infer the sexual reproduction rate, $F_{is}$ was estimated using Genetix (version 4.05; Belkhir et al., 2004). We also aimed to compare the results with the unbiased estimator of the complement of the Simpson index (estimator *L*) of heterogeneity, known as one of the indexes least sensitive to sample size (Lande, 1996) and modified to vary positively with heterogeneity (Pielou, 1969) between 0 and approximately 1-(1/G):

$$D* = 1 - L = 1 - \sum_{i=1}^{G} \left[ \frac{n_i(n_i - 1)}{N(N - 1)} \right]$$

where $n_i$ is the number of sampled units with $MLL_i$.

*R* and *D\** were computed for each population, and $F_{is}$ was computed for each population with (ramet level) and without (genet level) replicates.

The relationships between *R* or *D\** and the $F_{is}$ were considered at the intra- and inter-species levels. A Shapiro test was performed to infer the normality of the distribution; when this null hypothesis was rejected, the Spearman correlation index between *R* or *D\** and the $F_{is}$ was computed, and its significance was assessed using R statistical software.

## Results

*Clonal richness, R*

The mean genotypic index *R* increased gradually from *P. oceanica* to *C. nodosa*, *Z. marina* and *Z. noltii* (Table 1, Figure 1), with the variance (reflected by the maximal and interquartile ranges) of these values decreasing from *P. oceanica* to *Z. noltii*. Only *C. nodosa* slightly departed from this trend, with a mean *R* lower than that of *P. oceanica* and the greatest maximal ranges of *R* values among the seagrasses. For the Simpson index $D^*$ (Table 1; Figure S1), the values were constrained to a narrower range, and the ranking of values among the four species was thus not as informative, although there were clearly lower values for the longer-lived species than for *Zostera* sp.

*Inbreeding coefficient, $F_{is}$*

The $F_{is}$ values and their means progressively increased from *P. oceanica* to *Z. noltii* (Table 1; Figure 1). Only *Z. noltii* showed a slightly positive mean $F_{is}$. The interquartile values were nearly strictly negative for *P. oceanica* and *C. nodosa*, with increasingly positive $F_{is}$ values from *Z. marina* to *Z. noltii*. This progression of the $F_{is}$ interquartile values was even clearer when the datasets were analysed without replicates (Figure 1) than with replicates (Figure S1).

**Table 1:** Summary of the clonal growth features (Duarte, 1991) for each of the four seagrass species, the average compiled values of genotypic heterogeneity (D*) and diversity (R) and departure from Hardy-Weinberg equilibrium (Fis) and the Spearman correlations between the two genotypic indices and the genetic index (r2; *: p<0.05; **: p<0.01;***: <0.001), which were obtained from the dataset including all genotyped ramets (with replicates, Fis AR) and the dataset including only genets (without replicates, Fis OG).

| | rhizome turnover | shoot lifespan | R | D* | with replicates (ramet level) | | | without replicates (genet level) | | |
|---|---|---|---|---|---|---|---|---|---|---|
| | | | | | $F_{is\ AR}$ | correlation $R \sim F_{is\ Ar}$ | correlation $D^* \sim F_{is\ AR}$ | $F_{is\ OG}$ | correlation $R \sim F_{is\ OG}$ | correlation $D^* \sim F_{is\ OG}$ |
| *Posidonia oceanica* | 0.09 | 11.98 | 0.57 | 0.88 | -0.19 | 0.31*** | 0.41*** | -0..09 | 0.11* | 0.16** |
| *Cymodocea nodosa* | 0.14 | 2.4 | 0.46 | 0.77 | -0.24 | 0.64*** | 0.68*** | -0..03 | 0.13** | 0.12* |
| *Zostera marina* | 2.19 | 1.52 | 0.64 | 0.96 | -0.08 | 0.29* | 0.28* | -0..04 | 0.16* | 0 |
| *Zostera noltii* | | | 0.75 | 0.93 | 0.02 | 0.05 | 0.04 | 0..02 | 0.05 | 0.02 |

*: <0.05    **: 0.01    ***: <0.001

**Figure 1.** Boxplot showing the average departure from Hardy-Weinberg equilibrium ($F_{is}$, without replicates, see Figure S1 for the same results at the ramet level) and genotypic richness (*R*) values over all studied meadows for each of the four seagrass species: *Posidonia oceanica*, *Cymodocea nodosa*, *Zostera marina* and *Z. noltii*.

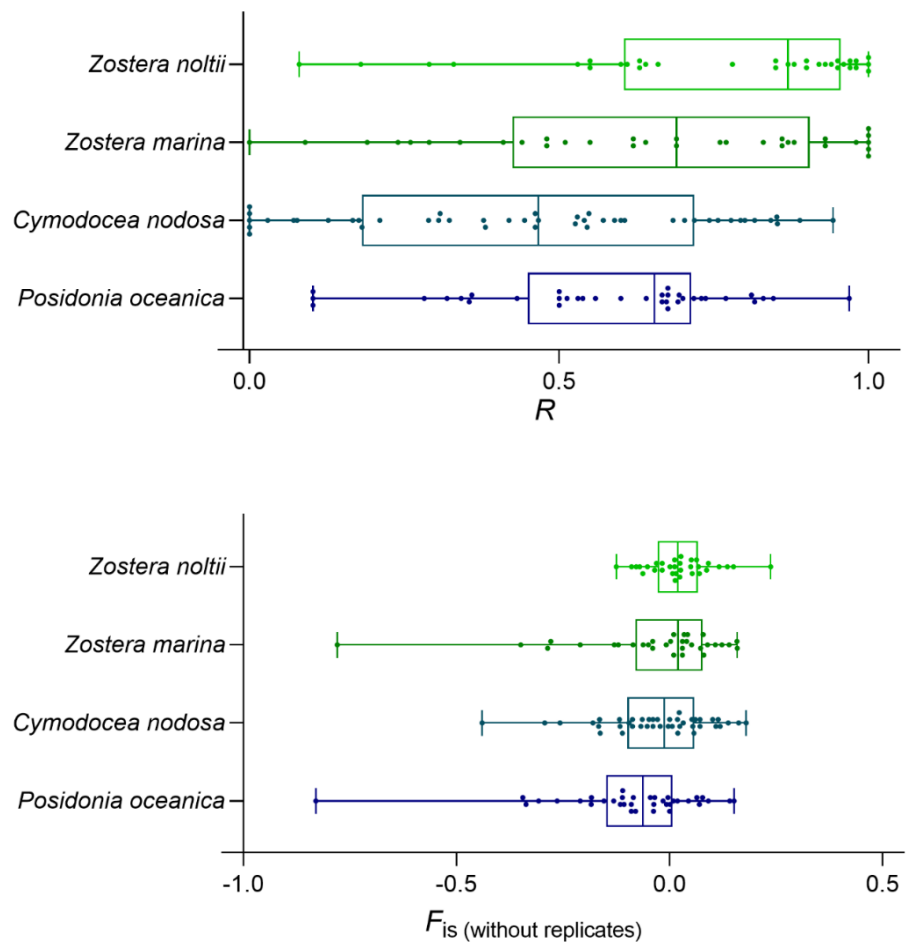

*Relationship between clonal indices and $F_{is}$*

At the intra-species level Table 1), the relationships between *R* or *D\** and $F_{is}$ were also positive and significant for all but the *Z. noltii* datasets. At the ramet level (Figure 2), the relationships ranged from highly positive (*P. oceanica* and *C. nodosa*) to slightly positive (*Z. marina*) and nearly null (*Z. noltii*). Similar results, though with weaker correlations, were obtained without replicates (Figure S2, Table 1), with the exception of a lack of significance for *Z. marina* at the genet level. Similar results were observed with the Simpson index of heterogeneity (*D\**; Figure S2b).

**Figure 2.** Relationships between the level of genotypic richness (R) and departure from Hardy-Weinberg equilibrium (Fis, at the ramet level, i.e., with replicates; see Figure S2 for the results at the genet level) at the meadow scale for each of the four seagrass species.

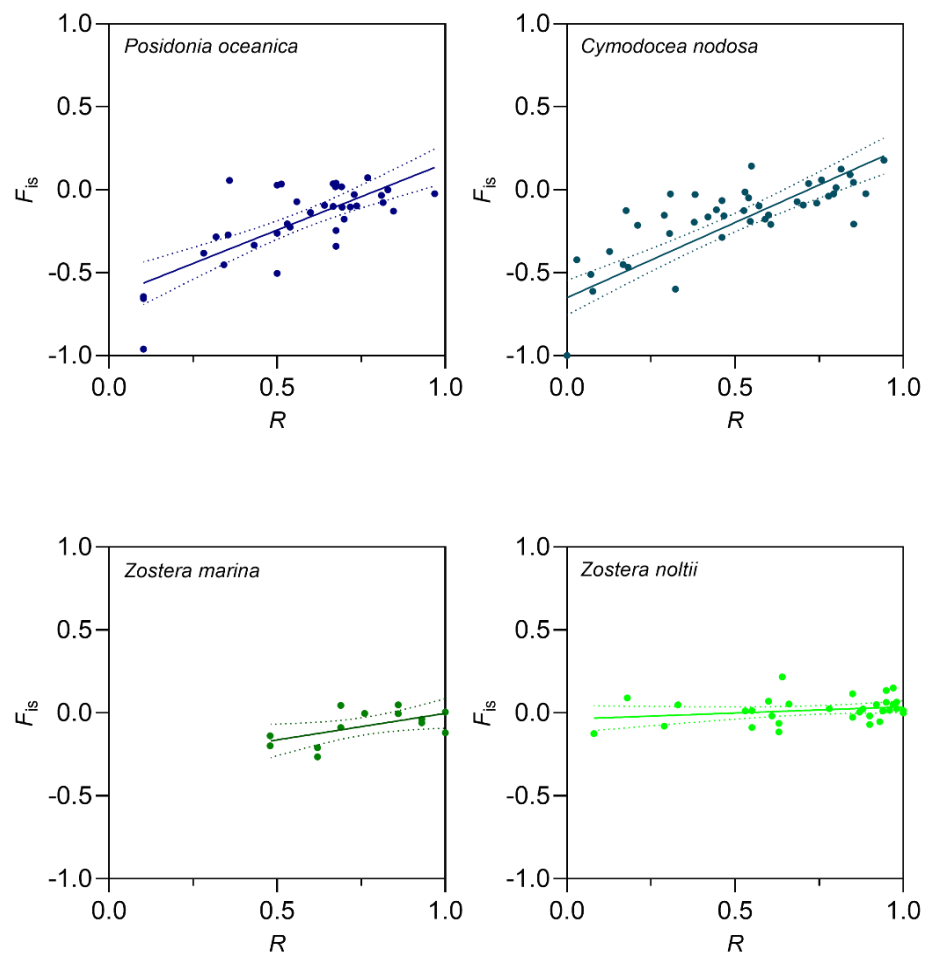

At the inter-species level, the relationships between $R$ and $F_{is}$ were positive and significant at the ramet level ($r^2$=0.41, p<0.0001; Figure 3). When considering only the genets (i.e., no replicates), the results were similar, but the correlations were weaker ($r^2$=0.12, p<0.0001, Figure S3). Similar relationships emerged between $D^*$ and $F_{is}$ at the ramet and genet levels ($r^2$=0.48, p<0.0001 and $r^2$=0.10, p<0.001, respectively; Figure 3 and S3).

**Figure 3:** Overall relationships between the genotypic richness (*R*) and Simpson index of heterogeneity (D*) and departure from Hardy-Weinberg equilibrium ($F_{is}$, at the ramet level, i.e., with replicates; see Figure S3 for the results at the genet level) at the meadow scale for each of the four seagrass species.

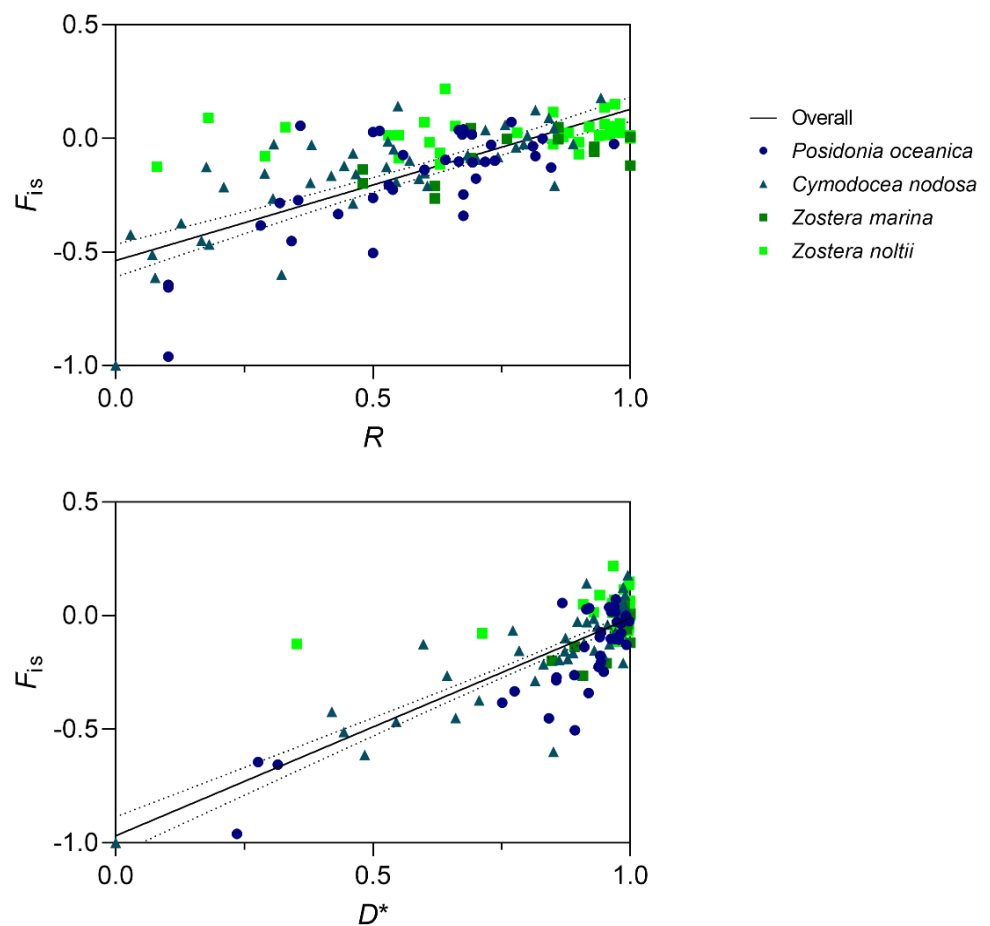

## Discussion

Here, we conducted a comparative re-analysis of genetic data from four species. The good ecological knowledge of the clonal architecture of these species allowed the testing and validation of predictions obtained from recent mathematical developments. Our analyses confirmed nearly all expectations derived from recent mathematical predictions about the relationships between 1) the turnover and longevity of shoots, 2) the prevalence of clonality, and 3) the consequential genotypic signatures and, above all, the genetic composition of natural meadows. When assessing the balance between clonal and sexual reproduction, by giving more weight to the genetic composition signature ($F_{is}$) than to the genotypic indices, which is highly sensitive to sampling (i.e., *R*), this analysis unravelled an apparent paradox reported in the literature: the frequent observation of apparently high genotypic richness, which suggests a high rate of sexual input, and systematic departure from HWE towards heterozygote excess, which suggests a very high rate of asexual reproduction. The results were in line with recent theoretical developments (Stoeckel et al., 2019), showing a barely discernible signature of *c* on *R* (also very weak on the Simpson index of heterogeneity *D**) and a much clearer signature on $F_{is}$. This trend was clear at both the inter-species and intra-species (among-meadow) levels, showing a nested pattern at both scales. Consequently, these results illustrate the need for a revised framework for the interpretation of molecular data from partial asexuals such as seagrasses, which may have important consequences for conservation and management strategies in the context of global change and increasing restoration needs (Carpenter et al., 2008; Hughes et al., 2008; Orth et al., 2006).

*Fitting empirical data to model predictions*

Our results support the ecological hypothesis that the importance of the clonal multiplication of ramets for the growth and maintenance of populations (compared to sexual input through seedlings) increases with increasing module lifespan.

First, the mean genotypic richness *R* gradually increases with decreasing longevity and increasing turnover of modules, from *P. oceanica* to *Z. noltii* (Figure 1), and the Simpson index of heterogeneity *D** follows a similar trend despite a narrower range of values. Although unable to provide a reliable quantitative assessment of the rates of clonality due to large subsampling bias (Arnaud-Haond et al., 2007b; Becheler et al., 2017; Gorospe et al., 2015; Stoeckel et al., 2019), *R* may remain useful in comparative studies and shows a broader range of discriminable scenarios than the Simpson index of heterogeneity *D**. With a comparable sampling scheme and effort, *R* (and *D**) can help assess the relative importance of clonality among natural populations. Such comparability was true here for *P. oceanica* and *C. nodosa* and partly for the *Zostera* species. It is thus important to bear in mind that the use of *R* (and *D**) here was by no means an attempt to estimate *c* but was for comparing the values of meadows sampled with similar sampling scales and strategies.

In fact, the high diversity of *Z. noltii* in this analysis may be influenced both by the short life span (and high turnover) of its shoots and the slightly different sampling designs used to gather the data analysed for the species. Indeed, the *Z. noltii* datasets were collected by sampling along mostly linear transects, while the datasets of the other three species were randomly or haphazardly sampled in a standardized area appropriated for their known clonal architecture, which should have minimized bias (Pielou, 1966) and enabled comparative analysis. Linear transects result in significantly higher genotypic diversities than other sampling methods due to the strong edge effect (Arnaud-Haond et al., 2007b). It is thus not possible to rigorously disentangle the effect of high rhizome turnover and low shoot longevity from the putative edge effects on the ranking of *Z. noltii* in terms of genotypic richness. However, the expected ranking for $F_{is}$ stands (Figure 1), suggesting life history traits had a stronger influence than sampling bias on these results.

This importance of sampling is reflected in a different way by the slight deviation from this general trend for *C. nodosa*. Despite having a slightly higher rhizome turnover than *P. oceanica* (Duarte, 1991a), *C. nodosa* shows lower mean *R* (as well as lower mean *D\**) values and a comparable distribution of $F_{is}$ values with even a tendency towards more extreme cases of heterozygote excess when all ramets are included (Figure S1). Several hypotheses can explain this result. First, despite a slightly higher rhizome turnover and a much shorter shoot life span in *C. nodosa* than in *P. oceanica*, the establishment and dynamics of *C. nodosa* meadows may rely more on clonal reproduction than those of *P. oceanica* meadows. Indeed, *C. nodosa* exhibits irregular sexual reproductive events dependent on environmental conditions but also exhibits a much faster clonal extension rate than *P. oceanica* (Marba et al., 1996; Marba et al., 2004). Moreover, although *C. nodosa* and *P. oceanica* were collected with an identical sampling strategy, the *C. nodosa* dataset contained more highly clonal meadows than the *P. oceanica* dataset, which may be partly due to its obligate dioecy (leading to colonizers, if alone, to rely on clonal growth), as well as the inclusion of sampling sites located at the limit of its distribution (Billingham et al., 2003). In fact, repeating the same analysis in Mediterranean populations delivered higher mean *R* (and *D\**) values, which were comparable only to those of *P. oceanica*, and a slightly less negative $F_{is}$, providing a better fit to the initial predictions.

Second, the increasing departure from HWE towards negative $F_{is}$ (heterozygote excess, Figure 1 and 2) from *Z. noltii* to *P. oceanica* also supports the prediction of mathematical models (Stoeckel et al., 2014). As the prevalence of clonality increases, the trajectory of the population towards equilibrium slows down for both positive and negative $F_{is}$ values, with less positive than negative $F_{is}$ values (Reichel et al., 2016). In fact, $F_{is}$ values are usually negative for *P. oceanica* and *C. nodosa*, and an increasing proportion of positive values are observed for *Z. marina* and *Z. noltii*. The tendency towards heterozygous excess for the first two species is even clearer when the data are considered at the genet (without replicates, Figure 1) rather than the ramet scale (Figure S1), with only interquartile $F_{is}$ values equal to or greater than 0.

Interestingly, this inter-species pattern is also observed at the nested, intra-species level (Figure 2 and Figure S2), showing that the relationship between estimates of clonal richness *R* and $F_{is}$ transcends the species boundary and applies within species at the meadow scale. This phenomenon is reflected by a positive correlation between $F_{is}$ and *R* among meadows for *P. oceanica*, *C. nodosa* and *Z. marina* (Table 1), again supporting the predictive power of population genetics models that explicitly take partial clonality into account. Using a standard sampling strategy, many more clonal meadows (i.e., those exhibiting lower $R$ values) tend to exhibit higher departures from HWE towards heterozygote excess (negative $F_{is}$ values), again in line with theoretical predictions.

Finally, three of the four species are hermaphrodites with no existing data suggesting self-incompatibility. Nevertheless, all of the species show heterozygote excess at the ramet level, and this excess is maintained at the genet level to a different extent for two of the species. Sexual reproduction significantly departing from random mating due to selfing or inbreeding could pull $F_{is}$ towards positive values reflecting heterozygote deficiency, eliminating at least part of the signature of clonality. Here, if any inbreeding or selfing occurred in populations of the three hermaphroditic species, negative $F_{is}$ values would still suggest a predominant influence of clonal over sexual reproduction. The occurrence of such departure to random mating may however explain the fact that *C. nodosa*, which is the only obligate outcrosser, is also the species that slightly departs from the global ranking expected among species by showing extreme heterozygote excess similar to or higher than the values obtained for the longer-lived *P. oceanica*.

Considering the accumulated ecological and physiological knowledge of the rhizome growth and dynamics of these species, our comparison of theoretical predictions, which come from a 'simple' Wright-Fisher-like model extended to explicitly include clonality, with the population genetics data gathered on four seagrass species, shows remarkable congruence. This result highlights the need i) to extend population genetics theory to predict the dynamics of genetic diversity while accounting for various reproductive systems and ii) to expand the production and interpretation of empirical data to enhance our understanding of the main drivers of demography and connectivity in natural populations, as well as the possibility of achieving these goals.

*Implications for understanding clonal versus sexual prevalence and its influence on the dynamics and evolution of natural populations*

Species ranges are the result of multiple ecological and evolutionary drivers, among which genetic drift, selection and dispersal are essential processes strongly influenced by both environmental factors and demography (Gaggiotti, 2017). The accurate use of molecular markers and population genetics tools and models thus requires a good theoretical understanding of the ways by which evolutionary forces, including the reproductive system, drive the temporal and spatial dynamics of genetic polymorphism to in turn infer demographic dynamics and history from molecular data. New theoretical works are needed not only to explore a broader range of life history traits (including, for example, the genetic architecture under the combined effect of clonality and haplo-diploidy or selfing) but also to target the development of estimators of *c* from genetic data. In fact, current indices based on classical ecological (i.e., non-exhaustive) sampling mostly provide imperfect proxies of the relative importance of clonal *versus* sexual reproduction (Stoeckel et al., 2019). Understanding the respective roles of clonal and sexual recruitment and dispersal in determining local demography and spatial connectivity is essential for forecasting the evolution of these features in the context of global change and future range shifts. This knowledge is also a prerequisite for defining accurate management measures, and the effort to obtain this knowledge has been a strong incentive underlying population genetic studies of seagrass during the past few decades (Alberto et al., 2005; Arnaud-Haond et al., 2007b; Arnaud-Haond et al., 2012; Kendrick et al., 2012). Although the importance of clonal growth in the colonization, expansion, and maintenance of meadows has long been acknowledged (Duarte, 1991b; Kendrick et al., 2012), recent studies have proposed migration *via* the production and dispersal of seeds as a possible central driver of the dynamics and persistence of seagrasses (Kendrick et al., 2012; McMahon et al., 2014). This hypothesis is rooted in the observation of moderate to elevated levels of genotypic and genetic diversity, together with limited genetic differentiation and isolation by distance, in a panel of species that contains the four species targeted here and the Australian *Posidonia australis*, the Pacific *Zostera pacifica* and the Atlantic *Thalassia testudinum* (Kendrick et al., 2012).

Despite being represented by a limited number of species (approximately 70), seagrasses are characterized by diverse life history traits (Hemminga Duarte, 2000; McMahon et al., 2014), as illustrated by the panel of four species examined here. The hypothesis proposed by Kendrick et al. (2012) may thus apply to some seagrass species, particularly short-lived species producing a large number of seeds (Phan et al., 2017), especially if one also accounts for the dispersal of vegetative fragments (McMahon et al., 2014). However, the prevalence of clonality is associated with a decrease in the loss of diversity due to the influence of drift (Reichel et al., 2016), which may partly explain the elevated levels of genetic diversity observed in well-established meadows. Those high levels of alpha diversity also intrinsically limit the maximum possible estimates of beta diversity (genetic differentiation as estimated through $F$st, Charlesworth, 1998; Gregorius, 2010; Hedrick, 1999b) and their saturation, which can be well described by the disruption of isolation-by-distance patterns over large scales (Kendrick et al., 2012).

Part of the observations that led to the hypothesis of the central role of seed dispersal may thus reflect the prevalence of clonal reproduction, regardless of whether high connectivity exists. In fact, similar observations of high diversity and limited differentiation with no large-scale pattern of isolation by distance led to the exploration of mutation patterns across the distribution range of the long-lived species *P. oceanica* and *C. nodosa* at the distribution range scale (Arnaud-Haond et al., 2014). Disentangling the spatial distribution pattern of polymorphism revealed the accumulation of somatic mutations through clonal propagation had a stronger influence than the pattern of dispersal of sexual propagules, implying the influence of clonal reproduction was stronger than that of sexual reproduction on the dynamics and evolution of meadows at large spatial (distribution range) and temporal (evolutionary) scales (Arnaud-Haond et al., 2014).

The results presented here thus support the co-existence of complex trade-offs between clonal and sexual reproduction and dispersal among seagrass species, which may be better understood by specifically taking into account the large diversity of life history traits these species express (McMahon et al., 2014). Disentangling these cases individually and accurately assessing the level of genotypic diversity, as well as the influence of genotypic diversity on the resistance of natural populations (Hughes et al., 2008; Massa et al., 2013; Reusch et al., 2005), are challenging but are also extremely important for designing sound management and restoration strategies in a fluctuating environment (McMahon et al., 2017), representing a major research axis to develop in the future. Marine connectivity research has been enriched in recent years by the improvement of predictive (oceanographic modelling, Robert Sponaugle, 2009) and molecular (next-generation sequencing and high-density genome scan) tools (Riginos et al., 2016; Selkoe et al., 2016). The former can be readily enriched through the extensive knowledge gained by seagrass ecologists on the broad diversity of life history traits determining the timing and extent of the production and dispersal of sexual and clonal propagules (Duarte, 1991a; McMahon et al., 2014). The latter will certainly benefit research on seagrass population genetics by allowing a finer-grained snapshot of the distribution of polymorphisms and the various origins of mutations at nested spatial scales to better disentangle the relative impact of different evolutionary forces on their emergence and spatial spread. Finally, the integration of oceanographic and molecular information in new integrative Bayesian frameworks of analysis is currently underway (Gaggiotti, 2017) and represents a promising path for grasping the diversity and complexity of seagrass strategies in terms of demography and dispersal.

## Data accessibility

## Acknowledgements

We wish to thank the consortium of the ANR Clonix projects funded by the French National Research Agency (projects CLONIX: ANR-11-BSV7-007 and Clonix2D ANR-18-CE32-0001) for very useful discussions and Jim Coyer for providing the *Z. noltii* dataset. We also wish to thank Olivier Hardy for recommending this work in Peer Community In, as well as Stacy A. Krueger-Hadfield, Ludwig Triest and an anonymous referee for their constructive comments on a former version of this manuscript. Version 5 of this preprint has been peer-reviewed and recommended by Peer Community In Evolutionary Biology (https://doi.org/10.24072/pci.evolbiol.100083).

# 1 Author Contributions

SAH and SS conceived the study, and SAH and DB wrote the manuscript. SAH and DB compiled the data and performed the meta-analysis, data exploration and interpretation. All authors contributed to editing. SAH and SS were responsible for securing funding. All authors have read and approved the final manuscript.

## Conflict of interest disclosure

The authors of this preprint declare that they have no financial conflict of interest with the content of this article. Sophie Arnaud-Haond is one of the PCI Ecology recommenders.

# References

Aitken, S. N., Yeaman, S., Holliday, J. A., Wang, T., & Curtis-McLane, S. (2008). Adaptation, migration or extirpation: climate change outcomes for tree populations. *Evolutionary Applications, 1*(1), 95-111. doi:doi:10.1111/j.1752-4571.2007.00013.x

Alberto, F., Gouveia, L., Arnaud-Haond, S., Perez-Llorens, J. L., Duarte, C. M., & Serrao, E. A. (2005). Within-population spatial genetic structure, neighbourhood size and clonal subrange in the seagrass Cymodocea nodosa. *Molecular Ecology, 14*(9), 2669-2681. doi:10.1111/j.1365-294X.2005.02640.x

Alberto, F., Massa, S., Diaz-Almela, E., Arnaud-Haond, S., Duarte, C. M., & Serrão, E. A. (2008). Genetic differentiation in the seagrass Cymodocea nodosa across the Mediterranean-Atlantic transition region. *Journal of Biogeography, 35*, 1279-1294.

Ali, S., Soubeyrand, S., Gladieux, P., Giraud, T., Leconte, M., Gautier, A., . . . Enjalbert, J. (2016). CloNcaSe: Estimation of sex frequency and effective population size by clonemate resampling in partially clonal organisms. *Molecular Ecology Resources, 16*(4), 845-861. doi:10.1111/1755-0998.12511

Arnaud-Haond, S., & Belkhir, K. (2007a). GENCLONE: a computer program to analyse genotypic data, test for clonality and describe spatial clonal organization. *Molecular Ecology Notes, 7*(1), 15-17.

Arnaud-Haond, S., Duarte, C. M., Alberto, F., & Serrão, E. A. (2007b). Standardizing methods to address clonality in population studies. *Molecular Ecology, 16*(24), 5115-5139.

Arnaud-Haond, S., Duarte, C. M., Diaz-Almela, E., Marbà, N., Sintes, T., & Serrão, E. A. (2012). Implications of Extreme Life Span in Clonal Organisms: Millenary Clones in Meadows of the Threatened Seagrass *Posidonia oceanica. PLoS ONE, 7*(2), e30454. doi:10.1371/journal.pone.0030454

Arnaud-Haond, S., Migliaccio, M., Diaz-Almela, E., Teixeira, S. J. L., Van De Vliet, M. S., Alberto, F., . . . Serrão, E. A. (2007c). Vicariance patterns in the Mediterranean Sea: east–west cleavage and low dispersal in the endemic seagrass Posidonia oceanica. *Journal of Biogeography, 34*(6), 963-976.

Arnaud-Haond, S., Moalic, Y., Hernández-García, E., Eguiluz, V. M., Alberto, F., Serrão, E. A., & Duarte, C. M. (2014). Disentangling the Influence of Mutation and Migration in Clonal Seagrasses Using the Genetic Diversity Spectrum for Microsatellites. *Journal of Heredity, 105*(4), 532-541. doi:10.1093/jhered/esu015.

Arnaud-Haond, S., Moalic, Y., Hernández-García, E., Eguiluz, V. M., Alberto, F., Serrão, E. A., & Duarte, C. M. (2014b). *Data from: Scaling of processes shaping the clonal dynamics and genetic mosaic of seagrasses through temporal genetic* *monitoring*. Retrieved from: https://doi.org/10.5061/dryad.3b8k6

Ayala, F. J. (1998). Is sex better? Parasites say "no". *Proceedings of the National Academy of Sciences of the United States of America, 95*(7), 3346-3348.

Bailleul, D., Stoeckel, S., & Arnaud-Haond, S. (2016). RClone: a package to identify MultiLocus Clonal Lineages and handle clonal data sets in r. *Methods in Ecology and Evolution, 7*(8), 966-970. doi:10.1111/2041-210X.12550

Becheler, R., Benkara, E., Moalic, Y., Hily, C., & Arnaud-Haond, S. (2013). *Data from: Scaling of processes shaping the clonal dynamics and genetic mosaic of seagrasses through temporal genetic monitoring*. Retrieved from: https://doi.org/10.5061/dryad.1vp70

Becheler, R., Benkara, E., Moalic, Y., Hily, C., & Arnaud-Haond, S. (2014). Scaling of processes shaping the clonal dynamics and genetic mosaic of seagrasses through temporal genetic monitoring. *Heredity, 112*(2), 114-121. doi:10.1038/hdy.2013.82

Becheler, R., Diekmann, O., Hily, C., Moalic, Y., & Arnaud-Haond, S. (2010). The concept of population in clonal organisms: mosaics of temporally colonized patches are forming highly diverse meadows of Zostera marina in Brittany. *Molecular Ecology, 19*(12), 2394-2407.

Becheler, R., Masson, J.-P., Arnaud-Haond, S., Halkett, F., Mariette, S., Guillemin, M.-L., . . . Stoeckel, S. (2017). ClonEstiMate, a Bayesian method for quantifying rates of clonality of populations genotyped at two-time steps. *Molecular Ecology Resources, 17*(6), e251-e267. doi:10.1111/1755-0998.12698.

Belkhir K., Borsa P., Chikhi L., Raufaste N. & Bonhomme F. 1996-2004 GENETIX 4.05, logiciel sous Windows TM pour la génétique des populations. Laboratoire Génome, Populations, Interactions, CNRS UMR 5000, Université de Montpellier II, Montpellier (France).

Billingham, M. R., Reusch, T. B. H., Alberto, F., & Serrão, E. A. (2003). Is asexual reproduction more important at geographical limits? A genetic study of the seagrass Zostera marina in the Ria Formosa, Portugal. *Marine Ecology Progress Series, 265*, 77-83. doi:10.3354/meps265077

Callaghan, T. V., Carlsson, B. Å., Jónsdóttir, I. S., Svensson, B. M., & Jonasson, S. (1992). Clonal Plants and Environmental Change: Introduction to the Proceedings and Summary. *Oikos, 63*(3), 341-347. doi:10.2307/3544959

Carlsson, B. Å., & Callaghan, T. V. (1994). Impact of climate change factors on the clonal sedge Carex bigelowii: implications for population growth and vegetative spread. *Ecography, 17*(4), 321-330. doi:doi:10.1111/j.1600-0587.1994.tb00109.x

Carpenter, K. E., Abrar, M., Aeby, G., Aronson, R. B., Banks, S., Bruckner, A., Wood, E. (2008). One-Third of Reef-Building Corals Face Elevated Extinction Risk from Climate Change and Local Impacts. *Science, 321*(5888), 560-563.

doi:10.1126/science.1159196

Charlesworth, B. (1998). Measures of divergence between populations and the effect of forces that reduce variability. *Molecular Biology and Evolution, 15*(5), 538-543. doi:10.1093/oxfordjournals.molbev.a025953

Cornelissen, J. H. C., Song, Y.-B., Yu, F.-H., & Dong, M. (2014). Plant traits and ecosystem effects of clonality: a new research agenda. *Annals of botany, 114*(2), 369-376. doi:10.1093/aob/mcu113

Coyer, J. A., Diekmann, O. E., Serrão, E., Procaccini, G., Milchakova, N., Pearson, G. A., Olsen, J. L. (2004). Population genetics of dwarf eelgrass Zostera noltii throughout its biogeographic range. *Marine Ecology Progress Series, 281*, 12.

de Meeûs, T., Prugnolle, F., & Agnew, P. (2007). Asexual reproduction: Genetics and evolutionary aspects. *Cellular and Molecular Life Sciences, 64*(11), 1355-1372. doi:10.1007/s00018-007-6515-2

den Hartog, C. (1970). The seagrasses of the world. *Internationale Revue der gesamten Hydrobiologie und Hydrographie, 56*(1), 275. doi:10.1002/iroh.19710560139

Diaz-Almela, E., Marbà, N., Alvarez, E., Santiago, R., Martinez, R., & Duarte, C. M. (2008). Patch dynamics of the Mediterranean seagrass Posidonia oceanica: Implications for recolonisation process. *Aquatic Botany, 89*(4), 397-403.

Diekmann, O. E., & Serrão, E. A. (2012a). *Data from: Range-edge genetic diversity: locally poor extant southern patches maintain a regionally diverse hotspot in the seagrass Zostera marina*. Retrieved from: https://doi.org/10.5061/dryad.2589rn16

Diekmann, O. E., & Serrão, E. A. (2012b). Range-edge genetic diversity: locally poor extant southern patches maintain a regionally diverse hotspot in the seagrass Zostera marina. *Molecular Ecology, 21*(7), 1647-1657. doi:10.1111/j.1365-294X.2012.05500.x

Duarte, C. M. (1991a). Allometric scaling of seagrass form and productivity. *Marine ecology progress series. Oldendorf, 77*(2), 289-300.

Duarte, C. M. (1991b). Seagrass depth limits. *Aquatic Botany, 40*(4), 363-377. doi:https://doi.org/10.1016/0304-3770(91)90081-F

Gaggiotti, O. E. (2017). *Metapopulations of Marine Species with Larval Dispersal: A Counterpoint to Ilkka's Glanville Fritillary Metapopulations* (Vol. 54): SPIE.

Gorospe, K. D., Donahue, M. J., & Karl, S. A. (2015). The importance of sampling design: spatial patterns and clonality in estimating the genetic diversity of coral reefs. *Marine Biology, 162*(5), 917-928. doi:10.1007/s00227-015-2634-8

Gregorius, H. R. (2010). Linking Diversity and Differentiation. *Diversity, 2*(3), 370.

Halkett, F., Simon, J.-C., & Balloux, F. (2005). Tackling the population genetics of clonal and partially clonal organisms. *Trends in Ecology & Evolution, 20*(4), 194-201.

Hedrick, P. W. (1999a). Perspective: Highly Variable Loci and their Interpretation in Evolution and Conservation. *Evolution, 53*, 313-318.

Hedrick, P. W. (2005). A standardized genetic differentiation measure. *Evolution, 59*(8), 1633-1638. doi:Doi 10.1554/05-076.1

Hedrick, W. (1999b). Perspective: highly variable loci and their interpretation in evolution and conservation. *Evolution, 53*(2), 313-318. doi:10.1111/j.1558-5646.1999.tb03767.x

Hemminga, M. A., & Duarte, C. M. (2000). *Seagrass ecology*. Cambridge University Press, Cambridge. http://dx.doi.org/10.1017/CBO9780511525551

Hughes, A. R., Inouye, B. D., Johnson, M. T. J., Underwood, N., & Vellend, M. (2008). Ecological consequences of genetic diversity. *Ecology Letters, 11*(6), 609-623. doi:doi:10.1111/j.1461-0248.2008.01179.x

Hughes, A. R., & Stachowicz, J. J. (2009). Ecological impacts of genotypic diversity in the clonal seagrass Zostera marina. *Ecology, 90*(5), 1412-1419. doi:10.1890/07-2030.1

Jost, L. (2008). G(ST) and its relatives do not measure differentiation. *Molecular Ecology, 17*(18), 4015-4026. doi:10.1111/j.1365-294X.2008.03887.x

Kendrick, G. A., Waycott, M., Carruthers, T. J. B., Cambridge, M. L., Hovey, R., Krauss, S. L., Verduin, J. J. (2012). The Central Role of Dispersal in the Maintenance and Persistence of Seagrass Populations. *BioScience, 62*(1), 56-65. doi:10.1525/bio.2012.62.1.10

Lande, R. (1996). Statistics and partitioning of species diversity, and similarity among multiple communities. *Oikos, 76*(1), 5-13. doi:Doi 10.2307/3545743

Marba, N., Cebrian, J., Enríquez, S., & Duarte, C. (1996). *Growth patterns of Western Mediterranean seagrasses: species-specific responses to seasonal forcing* (Vol. 133).

Marba, N., Duarte, C., Alexandre, A., & Cabaço, S. (2004). How do seagrasses grow and spread? In (pp. 11-18).

Marba, N., & Duarte, C. M. (1998). Rhizome elongation and seagrass clonal growth. *Marine Ecology Progress Series, 174*, 12.

Massa, S. I., Arnaud-Haond, S., Pearson, G. A., & Serrão, E. A. (2009). Temperature tolerance and survival of intertidal populations of the seagrass *Zostera noltii* (Hornemann) in Southern Europe (Ria Formosa, Portugal). *619*(1), 195-201. doi:10.1007/s10750-008-9609-4

Massa, S. I., Paulino, C. M., Serrao, E. A., Duarte, C. M., & Arnaud-Haond, S. (2013). Entangled effects of allelic and clonal (genotypic) richness in the resistance and resilience of experimental populations of the seagrass *Zostera noltii* to diatom invasion. *BMC Ecol, 13*, 39. doi:10.1186/1472-6785-13-39

McMahon, K., van Dijk, K.-j., Ruiz-Montoya, L., Kendrick, G. A., Krauss, S. L., Waycott, M., Duarte, C. M. (2014). The movement ecology of seagrasses. *Proceedings of the Royal Society B: Biological Sciences, 281*(1795), 20140878. doi:doi:10.1098/rspb.2014.0878

McMahon, K. M., Evans, R. D., van Dijk, K.-j., Hernawan, U., Kendrick, G. A., Lavery, P. S., Waycott, M. (2017). Disturbance Is an Important Driver of Clonal Richness in Tropical Seagrasses. *Frontiers in Plant Science, 8*(2026). doi:10.3389/fpls.2017.02026

Navascues, M., Stoeckel, S., & Mariette, S. (2010). Genetic diversity and fitness in small populations of partially asexual, self-incompatible plants. *Heredity, 104*(5), 482-492.

Orth, R. J., Carruthers, T. J. B., Dennison, W. C., Duarte, C. M., Fourqurean, J. W., Heck, K. L., Williams, S. L. (2006). A Global Crisis for Seagrass Ecosystems. *BioScience, 56*(12), 987-996. doi:10.1641/0006-3568(2006)56[987:AGCFSE]2.0.CO;2

Pecuchet, L., Reygondeau, G., Cheung, W. W. L., Licandro, P., van Denderen, P. D., Payne, M. R., & Lindegren, M. (2018). Spatial distribution of life-history traits and their response to environmental gradients across multiple marine taxa. *Ecosphere, 9*(10), e02460. doi:doi:10.1002/ecs2.2460

Phan, T. T. H., De Raeymaeker, M., Luong, Q. D., & Triest, L. (2017). Clonal and genetic diversity of the threatened seagrass Halophila beccarii in a tropical lagoon: Resilience through short distance dispersal. *Aquatic Botany, 142*, 96-104. doi:https://doi.org/10.1016/j.aquabot.2017.07.006

Pielou, E. C. (1966). The measurement of diversity in different types of biological collections. *Journal of Theoretical Biology, 13*(Supplement C), 131-144. doi:https://doi.org/10.1016/0022-5193(66)90013-0

Pielou, E. C. (1969). *An introduction to mathematical ecology*. New-York: Wiley-Interscience.

Polidoro, B. A., Carpenter, K. E., Collins, L., Duke, N. C., Ellison, A. M., Ellison, J. C., Yong, J. W. H. (2010). The Loss of Species: Mangrove Extinction Risk and Geographic Areas of Global Concern. *PLoS ONE, 5*(4), e10095. doi:10.1371/journal.pone.0010095

R Core Team. (2015). R: A language and environment for statistical computing R Foundation for Statistical Computing, Vienna, Austria. Retrieved from http://www.R-project.org/

Reichel, K., Masson, J.-P., Malrieu, F., Arnaud-Haond, S., & Stoeckel, S. (2016). Rare sex or out of reach equilibrium? The dynamics of F(IS) in partially clonal organisms. *BMC Genetics, 17*, 76. doi:10.1186/s12863-016-0388-z

Reusch, T. B. H., Ehlers, A., Hämmerli, A., & Worm, B. (2005). Ecosystem recovery after climatic extremes enhanced by genotypic diversity. *Proceedings of the National Academy of Sciences of the United States of America, 102*(8), 2826. doi:10.1073/pnas.0500008102

Rice, W. (1989). Analyzing tables of statistical tests. *Evolution, 43*, 223-225.

Riginos, C., Crandall, E. D., Liggins, L., Bongaerts, P., & Treml, E. A. (2016). Navigating the currents of seascape genomics: how spatial analyses can augment population genomic studies. *Current zoology, 62*(6), 581-601. doi:10.1093/cz/zow067

Robert, K. C., & Sponaugle, S. (2009). Larval Dispersal and Marine Population Connectivity. *Annual Review of Marine Science, 1*(1), 443-466. doi:10.1146/annurev.marine.010908.163757

Selkoe, K., D'Aloia, C., Crandall, E., Iacchei, M., Liggins, L., Puritz, J., Toonen, R. (2016). A decade of seascape genetics: contributions to basic and applied marine connectivity. *Marine Ecology Progress Series, 554*, 1-19. doi:10.3354/meps11792

Sintes, T., Marbà, N., & Duarte, C. (2006). Modeling nonlinear seagrass clonal growth: Assessing the efficiency of space occupation across the seagrass flora. *Estuaries and Coasts, 29*(1), 72-80. doi:10.1007/BF02784700

Stoeckel, S., & Masson, J.-P. (2014). The Exact Distributions of Fis under Partial Asexuality in Small Finite Populations with Mutation. *PLoS ONE, 9*(1), e85228.

Stoeckel, S., Porro, B., & Arnaud-Haond, S. (2019). Revising upward our appraisal of clonal rates in partially clonal organisms: the discernible and the hidden effects of clonality on the genotypic and genetic states of populations. *ArXiv:1902.09365 [q-Bio] v4 peer-reviewed and recommended by Peer Community in Evolutionary Biology.* Retrieved from *http://arxiv.org/abs/1902.09365v4*. doi: arXiv:1902.09365

Tibayrenc, M., Kjellberg, F., & Ayala, F. J. (1990). A clonal theory of parasitic protozoa: the population structures of Entamoeba, Giardia, Leishmania, Naegleria, Plasmodium, Trichomonas, and Trypanosoma and their medical and taxonomical consequences. *Proceedings of the National Academy of Sciences, 87*(7), 2414-2418.

Waycott, M., Duarte, C. M., Carruthers, T. J. B., Orth, R. J., Dennison, W. C., Olyarnik, S., Williams, S. L. (2009). Accelerating loss of seagrasses across the globe threatens coastal ecosystems. *Proceedings of the National Academy of Sciences, 106*(30), 12377-12381. doi:10.1073/pnas.0905620106

Yu, F.-H., Roiloa, S. R., & Alpert, P. (2016). Editorial: Global Change, Clonal Growth, and Biological Invasions by Plants. *Frontiers in Plant Science, 7*, 1467-1467. doi:10.3389/fpls.2016.01467

# Supplementary Material

**Figure S1.** Boxplot showing the average departure from Hardy-Weinberg equilibrium ($F_{is}$, with replicates) and genotypic richness (*R*) and heterogeneity (*D**) values over all studied meadows for each of the four seagrass species: *Posidonia oceanica*, *Cymodocea nodosa*, *Zostera marina* and *Z. noltii*.

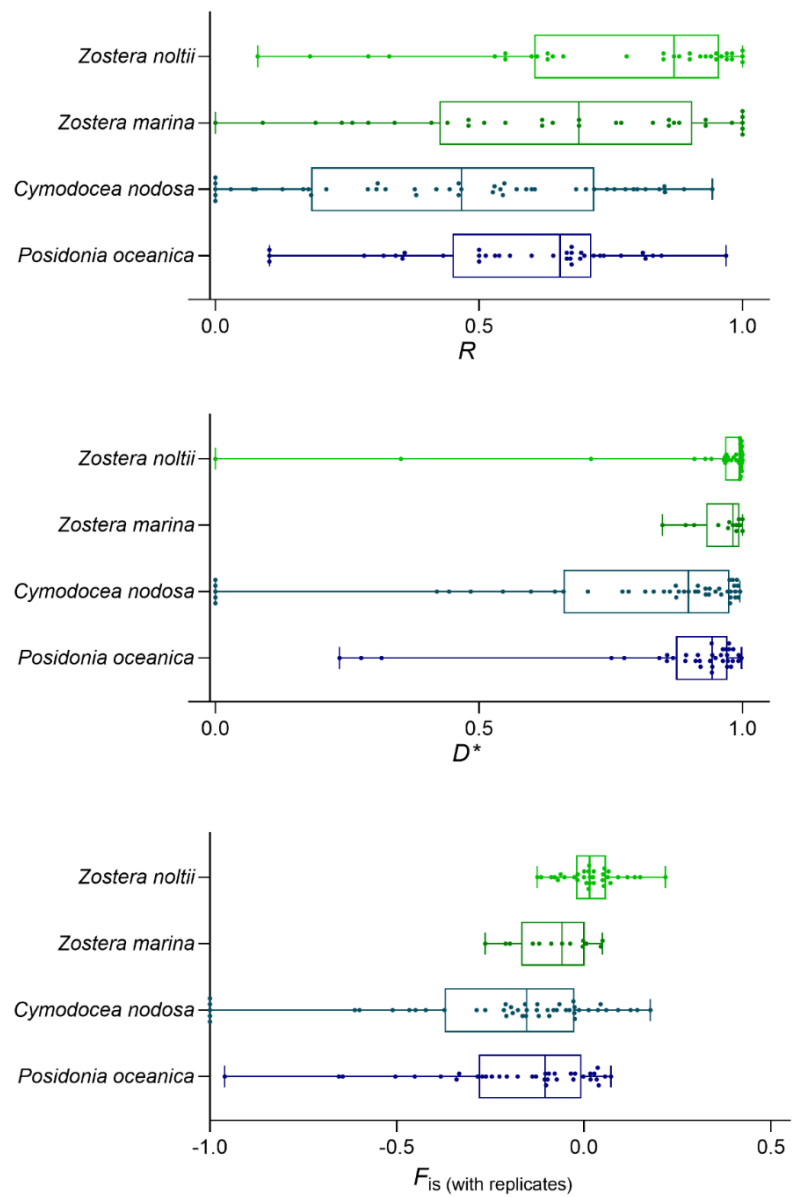

**Figure S2.** Relationships between the level of genotypic richness (*R*) and the departure from Hardy-Weinberg equilibrium ($F_{is}$, at the genet level, i.e., without replicates) at the meadow scale for each of the four seagrass species.

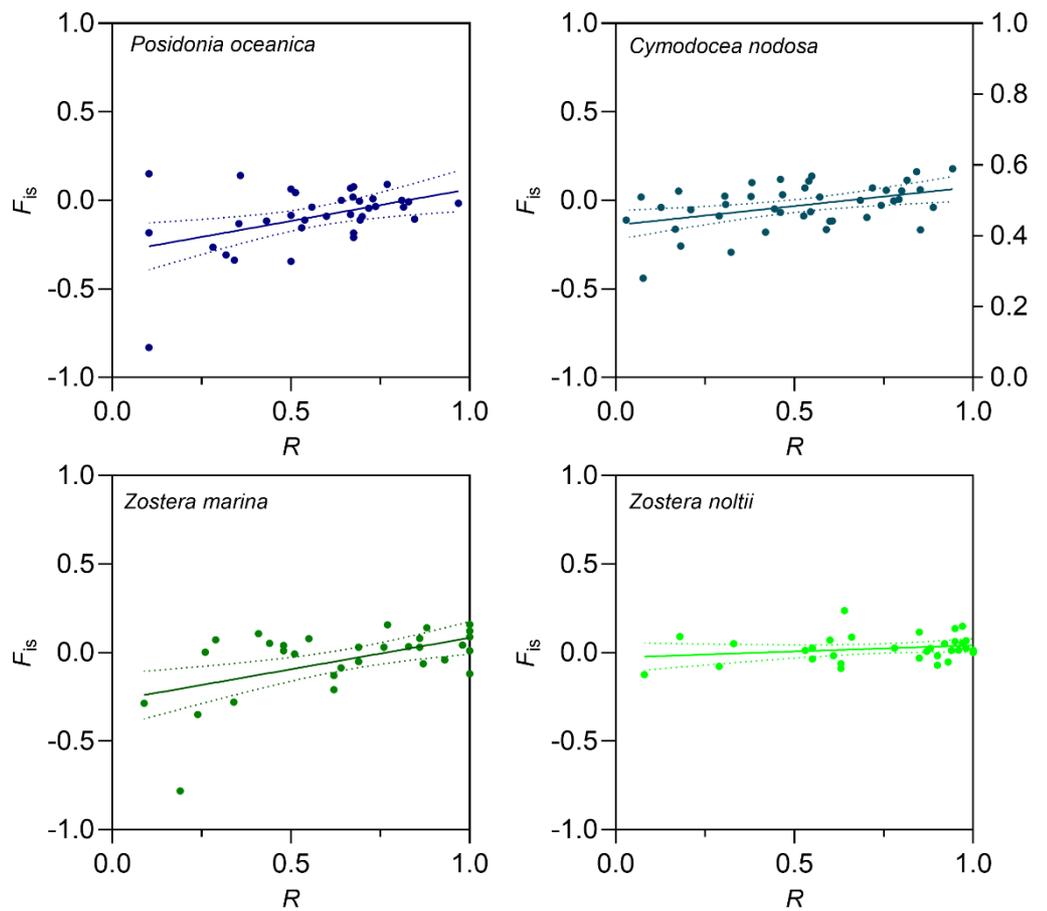

.

**Figure S2b.** Relationships between the Simpson index of heterogeneity ($D^*$) and the departure from Hardy-Weinberg equilibrium ($F_{is}$, at the ramet level, i.e., full datasets in the upper part, and at the genet level, i.e., without replicates, in the lower part) at the meadow scale for each of the four seagrass species.

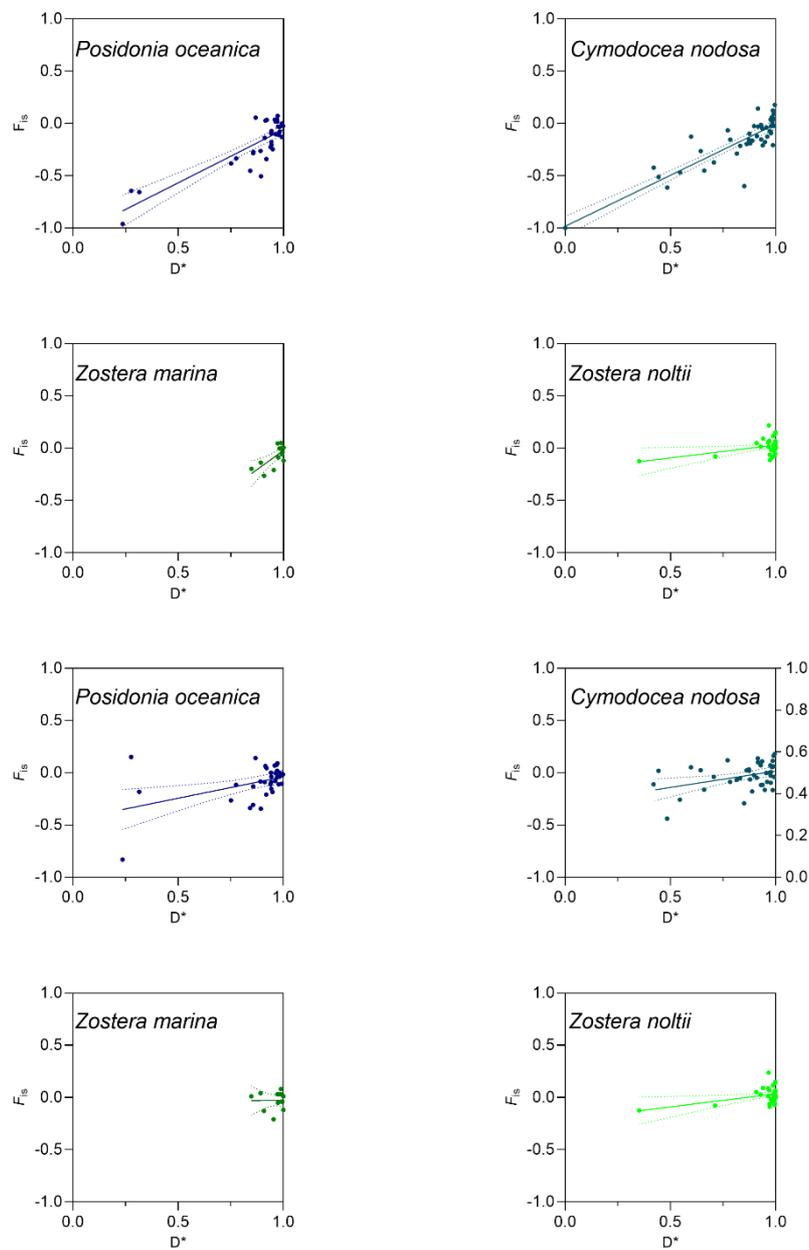

**Figure S3:** Overall relationships between the genotypic richness ($R$) Simpson index of heterogeneity ($D^*$), and departure from Hardy-Weinberg equilibrium ($F_{is}$, at the genet level, i.e., without replicates) at the meadow scale for each of the four seagrass species.

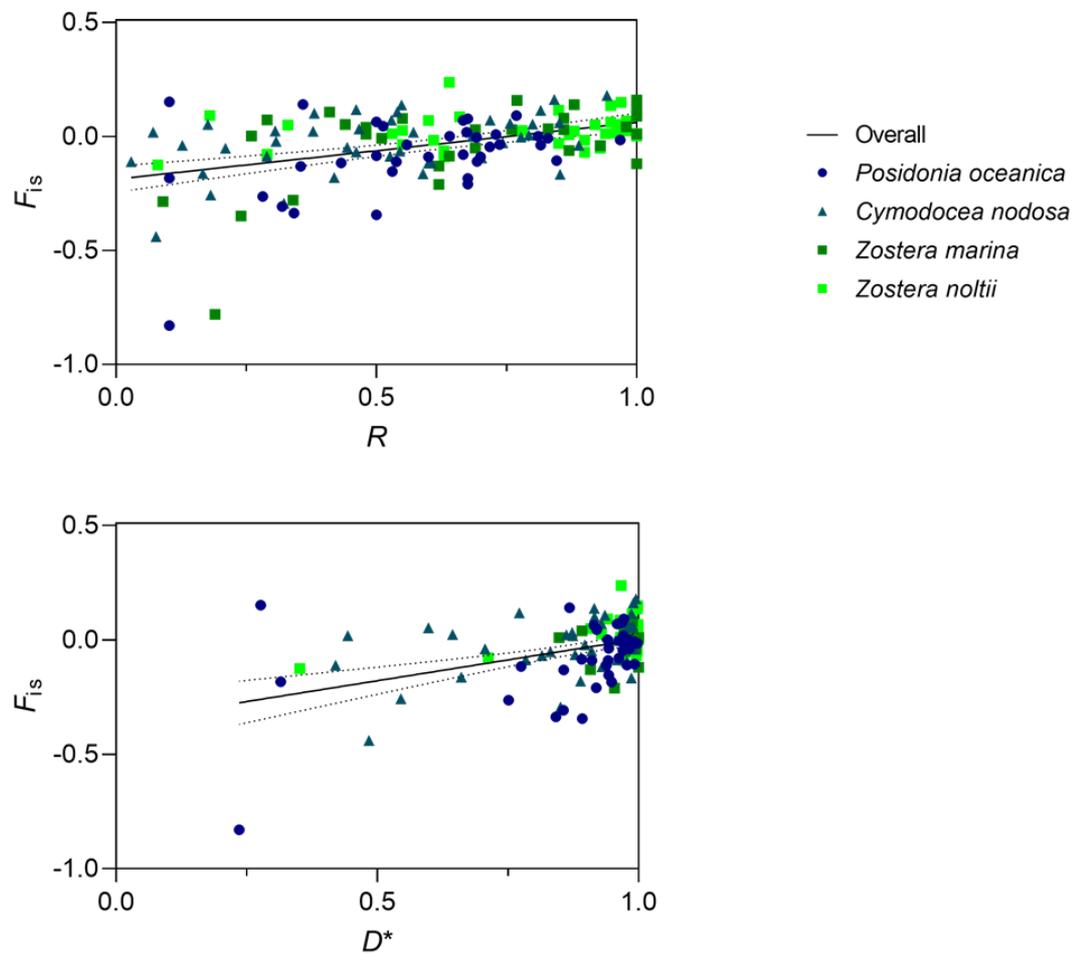